\documentclass[11pt,a4paper]{article}
\usepackage{authblk}
\usepackage{amssymb}
\usepackage{amsthm}

\usepackage[figuresright]{rotating}
\providecommand{\keywords}[1]{\textbf{\textit{Keywords:}} #1}

\begin{document}

\title{Experimental Study of Phase Transition in Pedestrian Flow}
\author[1]{Marek Buk\'a\v{c}ek}
\author[1,2]{P. Hrab\'ak}
\author[1]{M. Krb\'alek}
\affil[1]{Faculty of Nuclear Sciences and Physical Engineering, Czech Technical University in Prague, Trojanova 13, 120 00 Prague, Czech Republic}
\affil[2]{Institute of Information Theory and Automation, Academy of Sciences of the Czech Republic, Pod Vodarenskou vezi 4, 182 08 Prague, Czech Republic}
\maketitle

\begin{abstract}
The transition between low and high density phases is a typical feature of systems with social interactions. This contribution focuses on simple evacuation design of one room with one entrance and one exit; four passing-through experiments were organized and evaluated by means of automatic image processing. The phase of the system, determined by travel time and occupancy, is evaluated with respect to the inflow, a controlled boundary condition. Critical values of inflow and outflow were described with respect to the transition from low density to congested state. Moreover, microscopic analysis of travel time is provided.  \\ 
\end{abstract}

\keywords{
experiment passing through; boundary induced phase transition; travel time.
}


\section{Introduction}
\label{sec:int}

Different aspect of pedestrian behaviour have been studied last ten years. The coexistence of different phases of crowd motion (\cite{SeyPorSch2010LNCS}) and associated phase transition is studied mostly by means of fundamenthal diagrams, relations between density and flow, or velocity (\cite{SchSey2009PedBeh}, \cite{SchChoNis2010}). The pedestrian flow (or traffic in general) is mainly affected by so called bottleneck, a part of the facility, where is the wide corridor or room connected to some narrow part or exit (\cite{SeySteWinRupBolKli2009TGF}). There are other significant aspect of pedestrian motion: cross-sections (\cite{Krezt2014TGF}), T-junctions (\cite{BolZhaSeyStef2011IEEE}) or waiting zones (\cite{DavGeiMayPfaRoy2013TRC}).      

In this article we consider a simple rectangular room with one multiple entrance and one exit placed on the opposite wall. This room can be considered to be a single passing-through segment of a large complex facility (\cite{WasLub2013JCA}), e.g a corridor with bottleneck on the egress site. The common idea of all presented experiments is as follows: at the beginning the room is empty; after the initiation (start of the evacuation, end of the football match, etc.) pedestrians are entering the room in order to pass through. Let us further consider that the width of the exit is fixed and pedestrians are entering the room with given inflow rate $\alpha$ [pedestrians/second $=$ ped/s] which is the only free parameter of the system. 

By means of such experiments we aim to answer following questions:
\begin{enumerate}
	\item What is the mechanism of the transition from the free flow through the room to the congestion in front of the exit?
	\item How does the transition relates to the inflow rate $\alpha$?
	\item What is the saturation value $\alpha_{S}$?
	\item How does the saturation reflects in the macroscopic quantities as average occupation of the room and average travel-time through the room?
	\item How is the phase transition related to microscopic personal characteristic (e.g. pedestrian travel time, time headway at the doors)
\end{enumerate}

Related studies have been presented in \cite{BukHraKrb2014LNCS} and \cite{BukHraKrb2014TGF}. In this article improved method for automatic pedestrian detection is presented, which enables better and more reliable recording of microscopic quantities as travel time through the room or pedestrian velocity. Furthermore, every pedestrian can be identified by unique code, therefore it is possible to discuss their individual properties according to the motion inside the crowd.

The idea of studying such open room is inspired by the work \cite{EzaYanNis2013JCA}, where a phase transition from low to high density is studied by means of the Floor Field model in similar environment as the room setting discussed in this article. The phenomenon of boundary induced phase transition is often observed in interacting particle systems and similar agent-based simulations (\cite{BukHra2014LNCS}). By the boundary conditions we understand the narrow exit (60~cm in presented experiment) limiting the outflow and the variable inflow of pedestrians (1--2 [ped/s] in the experiments). In dependence on internal parameters as friction function parameter  $\zeta$, friction parameter $\mu$, used updating scheme etc., the system evinces a saturation of the outflow after crossing over certain ``saturation point'', or critical value $\alpha_S$ of the inflow $\alpha$. This leads to reaching the capacity of the corridor or room and results in creation of a growing cluster in front of the exit.

Such dependence on the inflow parameter is called phase transition in this article. It is important to note that there is another type of transition to be considered in such systems, and hence, the transition from the free flow at the beginning of the experiment or simulation to the congestion created after certain time of the system evolution. Such transition is not discussed in this document.

To compare the mechanism of the phase transition observed in simulations with the (semi)real system, four experiments were established at the Faculty of Nuclear Sciences and Physical Engineering (FNSPE) from 2012 to 2014. In the following, the experiments are indexed as E1, E2, E3, and E4. The first experiment E1 discussed in \cite{HraBukKrb2013JCA} concerned with the egress situation only. In this article the results of time-headway evolution at the exit have been compared with the experiment E4. Experiments E2 and E3 focused mainly on fundamental diagram evaluation; the phase transitions were evaluated from the macroscopic point of view mainly, i.e., the creation of the stable cluster or growing congestion has been observed, for detail see \cite{BukHraKrb2014LNCS} or \cite{BukHraKrb2014TGF}. The experiment E4 was designed better to capture the phase transition by the change of the travel time through the room as well. The experimental setting is in detail described in section~\ref{sec:setting}.

\section{Experimental setting}
\label{sec:setting}

\begin{figure}[h!]
	\begin{center}
	\hfill\includegraphics[height=.25\textwidth]{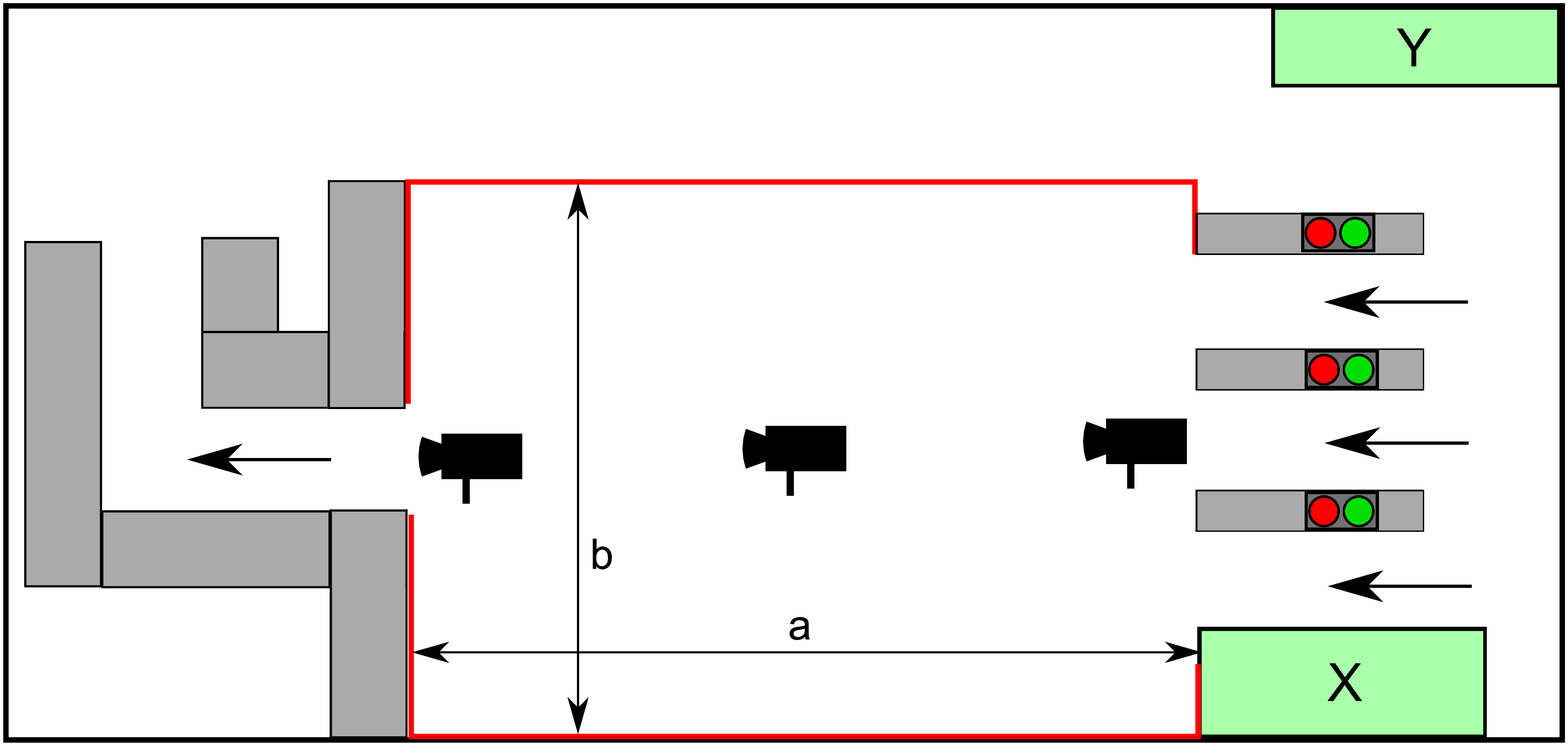}\hfill
	\includegraphics[height=.25\textwidth]{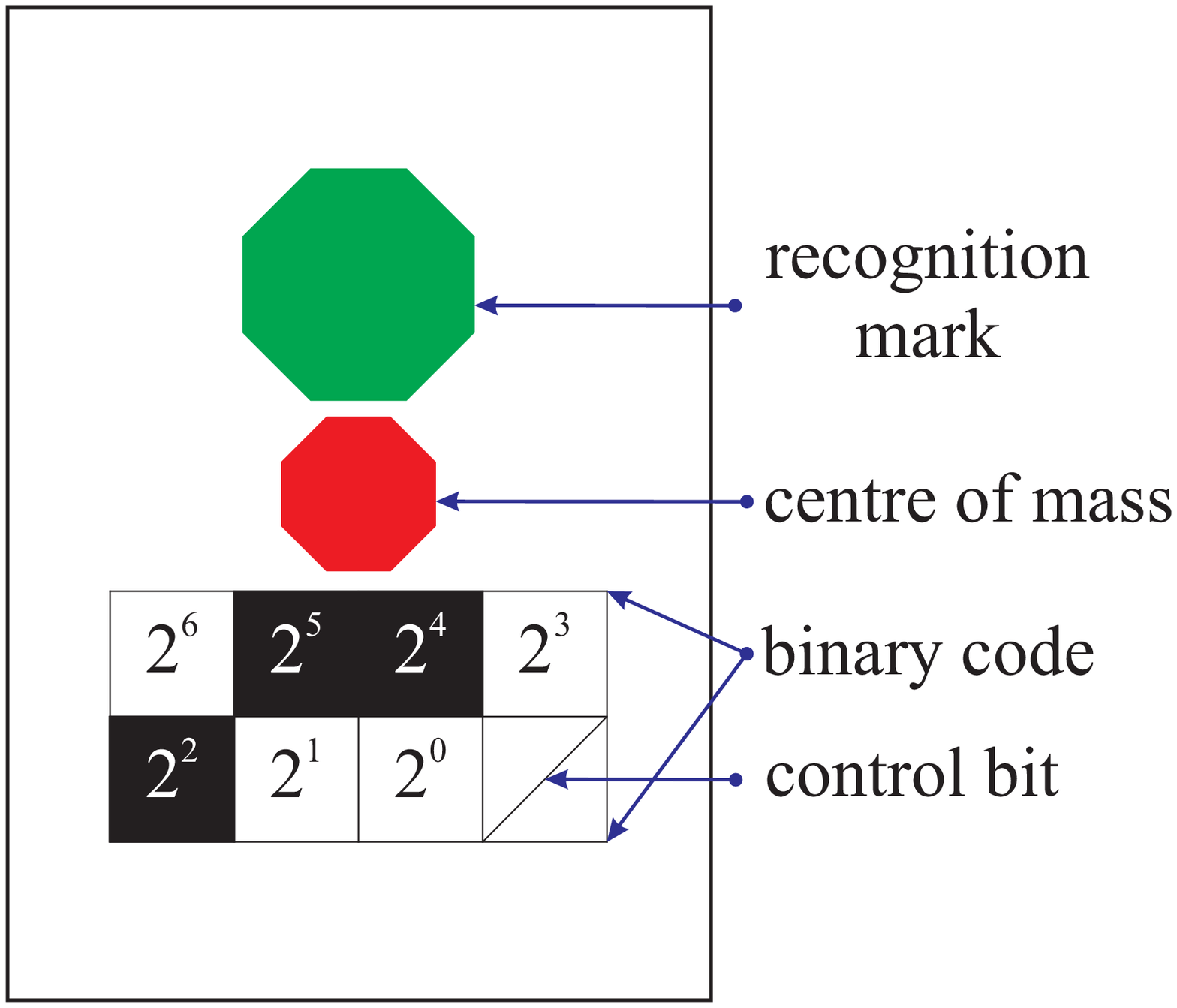}\hfill\phantom{x}
	\end{center}
\caption{Left: Experimental setting of E2, E3, and E4. The dimensions of experimental room are as follows: E4: a = 8 m, b = 4.5 m, E3: a = 7.2 m, b = 3.9 m, E2: a = 7.8 m, b = 5.5 m. Black circles represent the position of cameras, technical support was situated in area X, area Y represents refreshment corner. Right: sketch of pedestrian's cap, its size corresponds to A4 sheet.}
\label{fig:setting}
\end{figure}

The setting of experiments E2, E3, and E4 is depicted in Figure~\ref{fig:setting}. Around 80 volunteers (second year students of FNSPE) were gathered in front of the exit. After initiation, the volunteers were instructed to enter the room at the green signal near the entrance and leave the room as fast as possible by walking (running and strong physical contacts were forbidden). 

\begin{figure}[h!]
	\begin{center}
	\includegraphics[width=.9\textwidth]{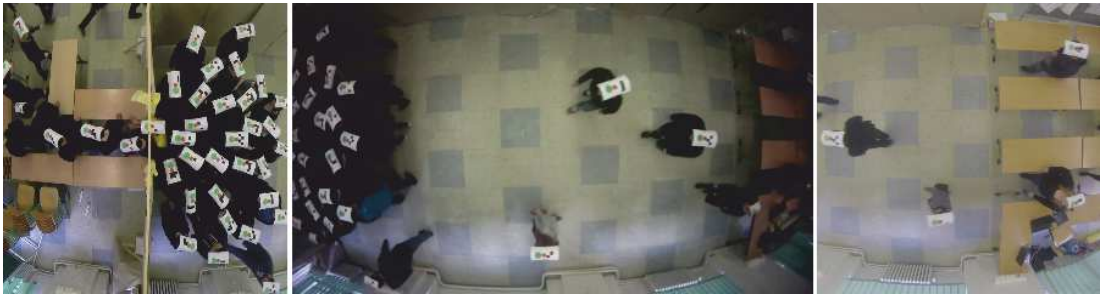}
	\end{center}
\caption{Snapshot from all three cameras at 2 min 21 s of the experiment E4, run \#7 ($\alpha=1.95$). Fluctuating cluster of $\approx 25$ pedestrians is visible.}
\label{fig:snapshot}
\end{figure}

The observed room was equipped by three cameras: above the center of the room, above the entrance, and above the exit. During the experiment E4 emphasis has been put on automatic detection of pedestrians. Every pedestrian was equipped by a hat with unique binary code, as depicted in Figure~\ref{fig:setting}. The pedestrian detection was performed by means of specific color recognition in each frame of the record. A semi-automatic software was used for the identification of the input and output of each pedestrian. Based on the manual comparison with the records in dens regime, the reliability of such detection is above 95 \%. A snapshot from the experiment is presented in Figure~\ref{fig:snapshot}.

To control the inflow rate $\alpha$, the signalling device was used. Randomly long red signal was alternated by 0.1 s green signal, on which, certain number of pedestrians were forced to enter. In experiments E2 and E3, two or three pedestrians entered the room together. Based on the experiment E2 it has been observed that the ability of pedestrians to react on the green signal is limited and the behaviour of pedestrian, who entered in a group, lacks their individual properties. Therefore, in E4 three independent devices, each at one part of the multiple entrance, have been used to support the randomness of the arrivals.

To simulate the independent arrival of pedestrians to the entrance, the geometric distribution for time intervals between two green signals has been used, i.e., the probability of the time interval between two signals on one device to be $\Delta t=k\cdot h$ is
\begin{equation}
	\Pr(\Delta t =k\cdot h)=(\alpha\cdot h/3)(1-\alpha\cdot h/3)^{k-1}\,,
\end{equation}
where $h=0.6$ is the shortest time interval on which the pedestrians can react relatively reliably (measured in pre-experiment, the lower value the more random arrivals). Since the expected value of such distribution is $3/\alpha$ seconds between two entrances of pedestrians through one entrance and considering that there are three entrances (controlled by independent signal devices), we have setted the inflow rate to $\alpha$.

According to the observations, real inflow of pedestrians was still little bit lower (around 90~\% of $\alpha$). Therefore, the comparison and data evaluation has been done with respect to the real inflow $\alpha_{\rm real}$.

\section{Inflow-Rate-Induced Phase Transition}
\label{sec:phtr}

The first insight in the mechanism of the phase transition gives the Table~\ref{tab:transition}. Individual runs of the experiment can be divided into three groups according to the macroscopic observation:
\begin{enumerate}
	\item \emph{Free flow}: no cluster is created in front of the exit. The interaction between pedestrians leads only to occasional delay at the exit but has no influence on the whole system
	\item \emph{Congestion}: the interaction in front of the exit blocks partially the possibility of pedestrians to leave the room in the same rate as they are entering it. This causes the creation of a cluster in front of the exit, which size grows and would fill the whole room (the experiment was stopped before this happened because of the lack of volunteers).
	\item \emph{Stable cluster}: conflicts near the exit lead to the creation of a stable cluster. The size of the cluster fluctuates between 15 and 30 pedestrians. This is the evidence that the inflow is close to the capacity of the exit.
\end{enumerate}

\begin{table}[h!]
\caption{The macroscopic evaluation of the phase transition from experiments E2, E3, and E4. $\alpha$ denotes the parameter of the signalling device at the input, $\alpha_{\rm real}$ is the real inflow obtained from data analyses, $J_{\rm out}$ is the measured outflow through the exit, $\overline{TT}$ the average travel time, $N_{150}$ the number of pedestrians in the room 150~s after the initiation.}
\label{tab:transition}  
	\begin{center}
	\hfill
	\begin{tabular}{p{.7cm}p{1cm}p{1cm}p{2.5cm}}
	\hline
	\multicolumn{4}{l}{Experiment E2}\\
	ID&$\alpha$ [ped/s]& input [ped]& observation\\
	\hline\hline
	\# 1&1.12&2&free flow\\
	\# 2&1.19&2&free flow\\
	\# 3&1.26&2&free flow\\
	\# 4&1.40&2&stable cluster\\
	\# 5&1.62&3&stable cluster\\
	\# 6&1.74&3&congestion\\
	\# 7&1.78&3&congestion\\
	\# 8&1.81&3&congestion\\
	\# 9&1.91&3&congestion\\
	\hline
	\end{tabular}\hfill
	\begin{tabular}{p{1.3cm}p{1cm}p{1cm}p{2.5cm}}
	\hline
	\multicolumn{4}{l}{Experiment E3}\\
	ID&$\alpha$ [ped/s]& input [ped]& observation\\
	\hline\hline
	\# 1&1.24&2&free flow\\
	\# 2&1.25&2&free flow\\
	\# 3&1.33&2&free flow\\
	\# 4&1.46&2&stable cluster\\
	\# 5&1.70&3&stable cluster\\
	\# 6&1.76&3&stable cluster\\
	\# 7&1.92&3&transition\\
	\# 8&1.94&3&transition\\
	\# 9-11&$>2$&2-3&congestion\\
	\hline
	\end{tabular}
	\hfill\phantom{x}\\[.4cm]
	
	\begin{tabular}{p{1cm}p{1cm}p{1cm}p{1cm}p{1cm}p{1cm}p{2.5cm}}
	\hline
	\multicolumn{7}{l}{Experiment E4}\\
	ID&$\alpha$ [ped/s]&$\alpha_{\rm real}$ [ped/s]& $J_{\rm out}$ [ped/s]&$\overline{TT}$\phantom{xx} [s]&$N_{150}$ [ped]& observation\\
	\hline\hline
	\#  2&1.20&0.99&0.99& \hspace{1.1ex}5.67&  \hspace{1.1ex}3&free flow\\
	\#  5&1.35&1.22&1.20& \hspace{1.1ex}6.73&  \hspace{1.1ex}7&free flow\\
	\#  4&1.50&1.37&1.30&16.59& 24&stable cluster\\
	\#  3&1.50&1.43&1.33&14.39& 22&stable cluster\\
	\#  6&1.65&1.39&1.31&20.40& 33&stable cluster\\
	\#  7&1.95&1.55&1.37&25.78& 45&transition\\
	\# 11&1.94&1.61&1.38&21.65&41&transition\\
	\#  9&2.25&1.78&1.37&24.06&47*&congestion\\
	\#  8&2.25&1.79&1.38&25.03&46*&congestion\\
	\# 10&2.40&1.78&1.37&23.33&44*&congestion\\
	\hline
	\end{tabular}
	\end{center}
\end{table}

Four runs of experiments from Table~\ref{tab:transition} are marked with the term ``transition''. In those cases (\#7 and \#8 of E3 and \#7 and \#11 of E4) it is not sure, whether the size of the cluster stabilized or whether there was a growing congestion. The runs have been stopped too early to decide. The reason for stopping the experiment was that there were too few pedestrians outside the room, so they were not able to react correctly to the green signal to maintain given inflow.

\begin{figure}[h!]
	\begin{center}
	\includegraphics[width=.95\textwidth]{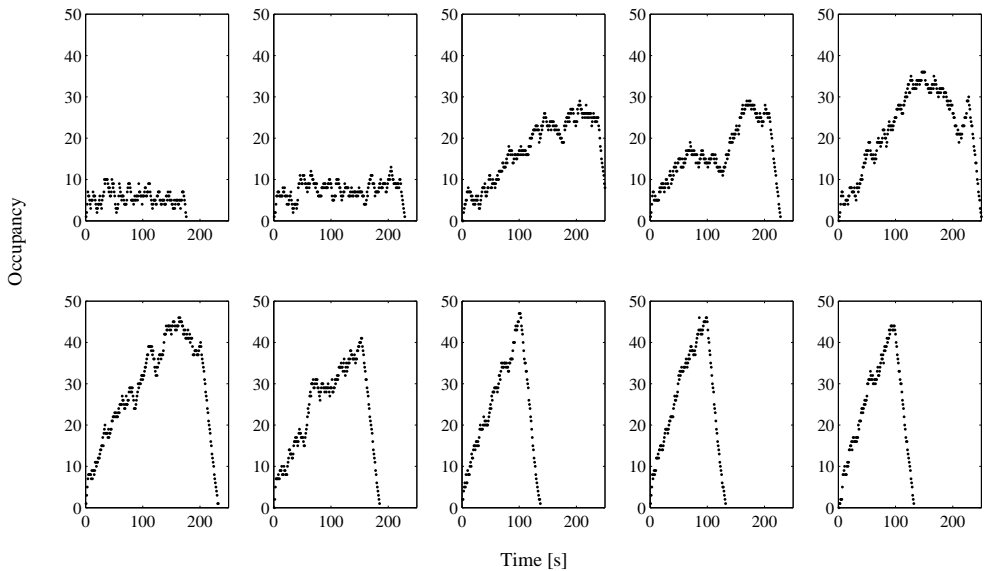}
	\end{center}
\caption{Number of pedestrians in the room plotted against the time, data taken from experiment E4, runs are ordered with respect to the inflow as in Table~\ref{tab:transition}.}
\label{fig:numofped}
\end{figure}

Similar situation occurred in runs \#8-10 of the experiment E4. The number of pedestrians denoted by the asterisk * is related to the maximal number of pedestrians in the room, which has been reached before 150~s from initiation; for detailed evolution of the number of pedestrians in the room see Figure~\ref{fig:numofped}.

From the above mentioned observation we can conclude that the saturation value of the inflow is between 1.40 and 1.60 ped/s (real inflow values from E2 and E3 have not been measured, given values are parameters of signalling device). Here we note that in the congestion phase there has not been observed the equivalent to the steady state. Nevertheless, the trend of the growth indicates that after longer time the room would have became completely filled by pedestrians.

The above mentioned conclusion is supported by the graphs in Figure~\ref{fig:numofped}, where the absolute occupation of the room is plotted against the time.

\begin{figure}[h!]
	\begin{center}
	\includegraphics[width=.45\textwidth]{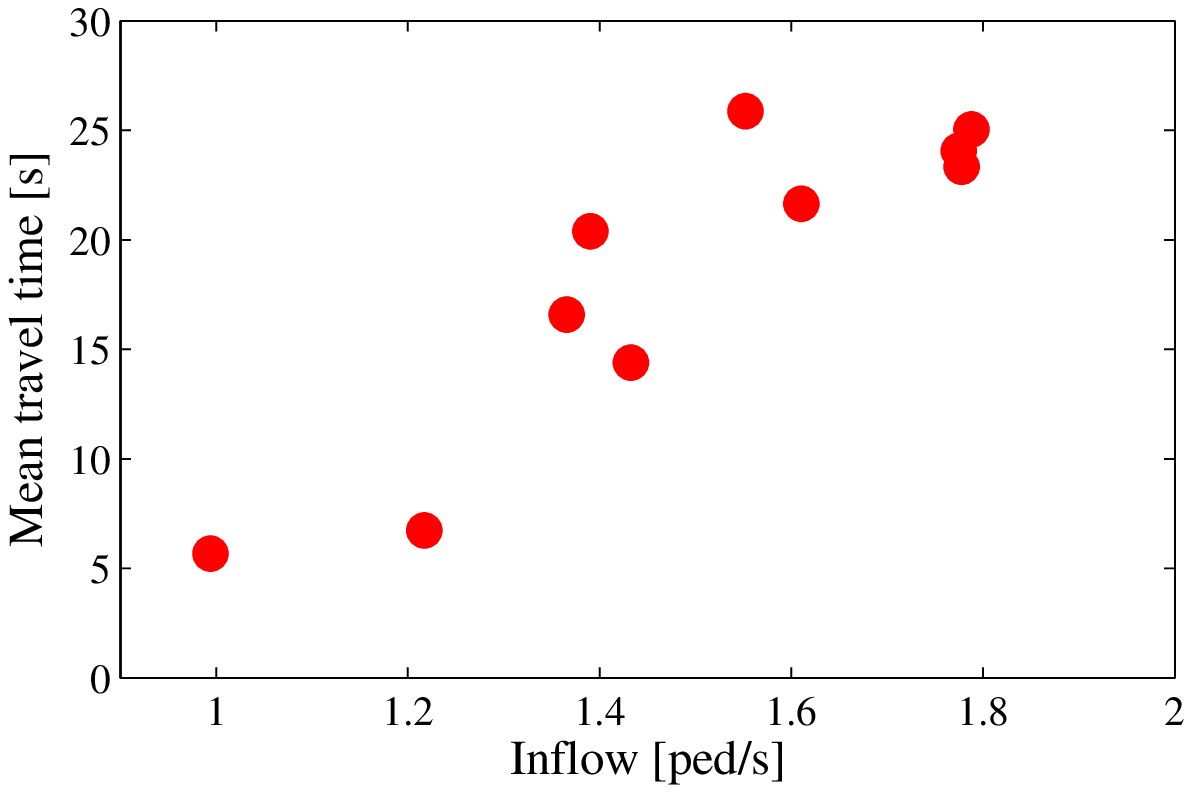}
	\includegraphics[width=.45\textwidth]{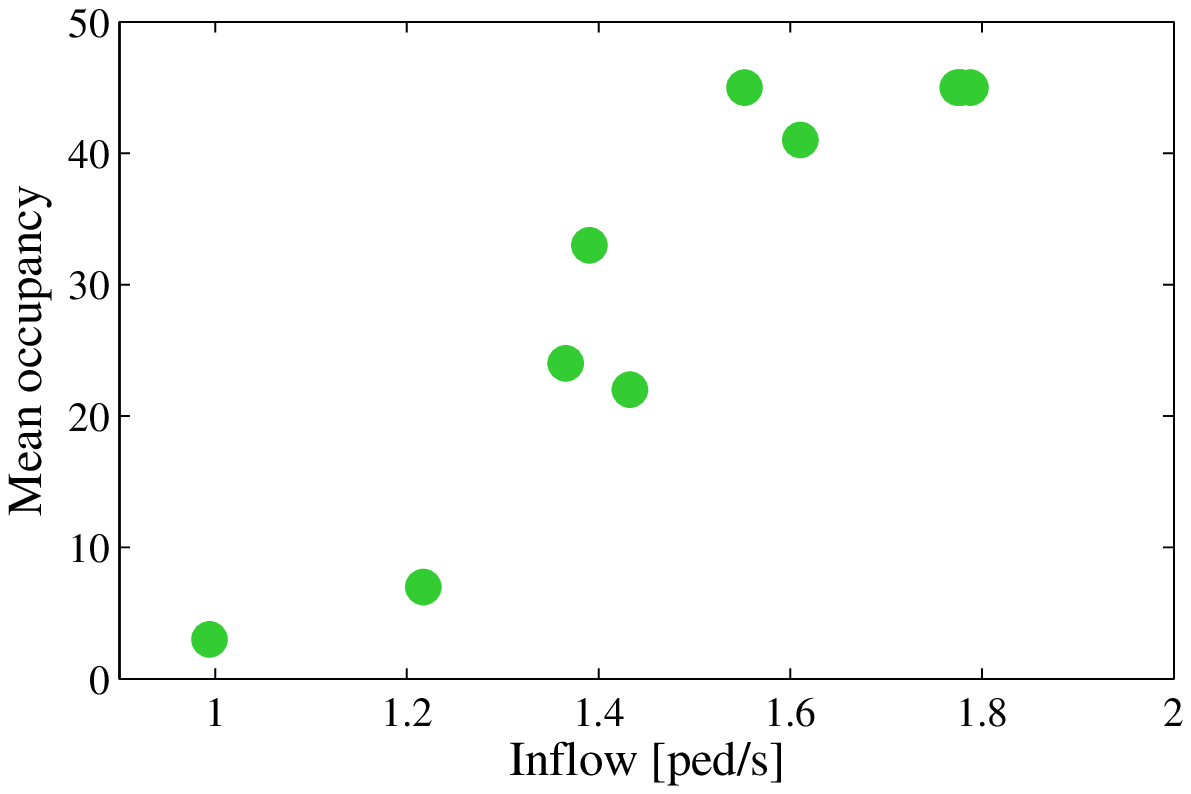}
	\end{center}
\caption{The relation of mean travel time (Left) and mean ocupancy (Right) to the inflow $\alpha_{\rm real}$, data taken from experiment E4.}
\label{fig:tt}
\end{figure}

The phase transition can be identified by the relation between the occupancy (measured in given time) and the inflow $\alpha_{\rm real}$, see Figure~\ref{fig:tt}. From the microscopic point of view, similar dependency describes the increase of the average travel time through the room. The travel time for each pedestrian was derived from the time of the entrance and the exit of the pedestrian. 

\begin{figure}[h!]
	\begin{center}
	\includegraphics[width=.45\textwidth]{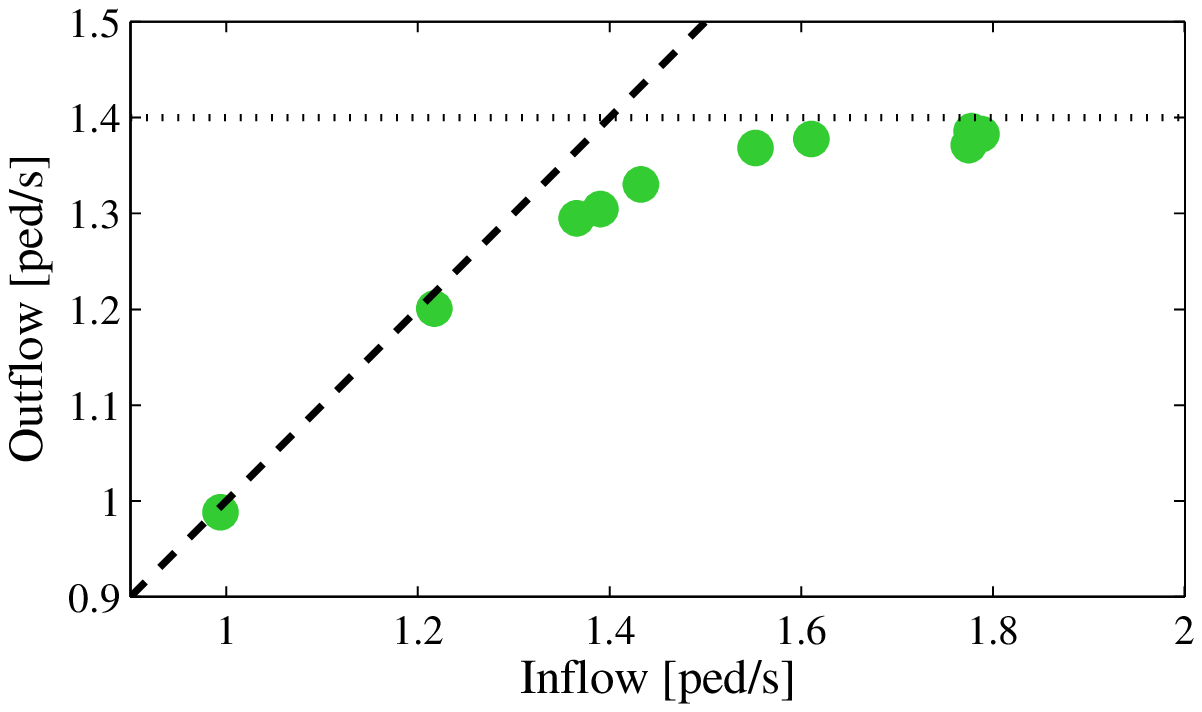}
	\includegraphics[width=.45\textwidth]{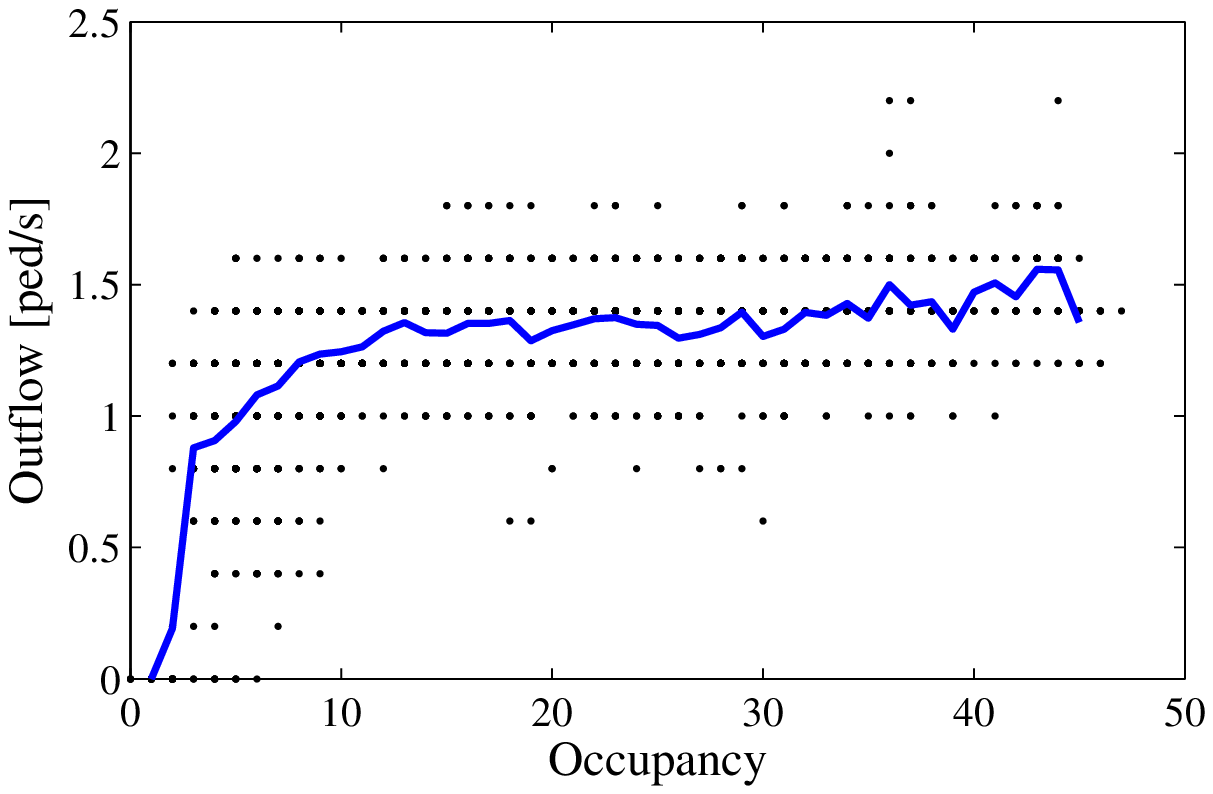}
	\end{center}
\caption{The dependency of outflow to inflow (Left) or occupancy (Right), data taken from experiment E4.}
\label{fig:out}
\end{figure}

The multiple entrance allowed us to increase the inflow of pedestrians above the limit of the outflow through the exit. The capacity of the exit (referred to as $J_{\rm MAX}$) during the experiment E4 was 1.38~ped/s, as visualized in Figure~\ref{fig:out}. This supports the hypothesis that realizations E4 \#7 and \#11 are very close to the phase transition. Therefore we conclude that the saturation point of the experimental setting is very close to $\alpha_{S}=1.4$~[ped/s] or the real inflow.

This trend corresponds to the dependency of the outflow on the number of pedestrians inside the room, see Figure~\ref{fig:out}. One can see that the outflow fluctuates around the value 1.38~ped/s after creation of small cluster in front of the exit. It is interesting that even in the huge cluster the conflicts do not block the movement, contrarily the outflow is little bit higher.

\section{Microscopical View}
\label{sec:micro}

Microscopic view is based mainly on the analysis of travel time. This quantity is affected by many factors (e.g desired velocity, willingness to overtake or to go through the dense crowd) and is significantly differers across pedestrians. 

\begin{figure}[h!]
	\begin{center}
	\includegraphics[width=.9\textwidth]{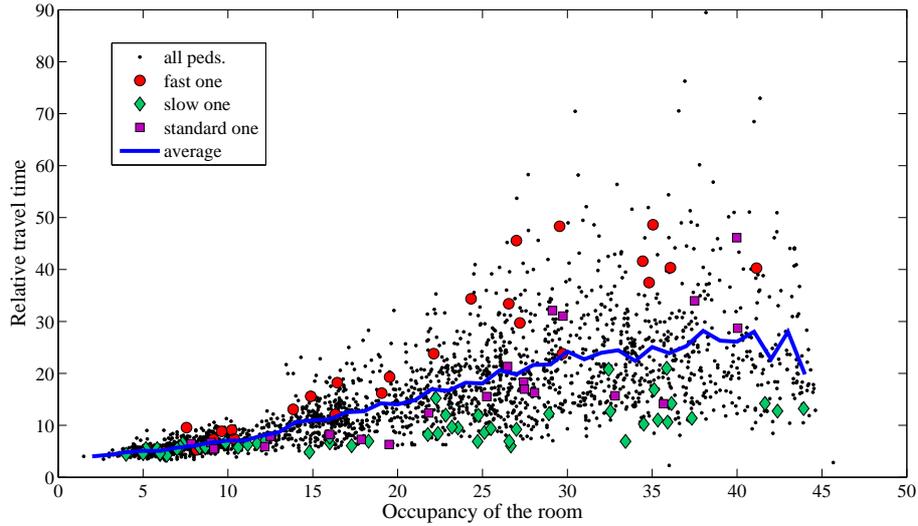}
	\end{center}
\caption{The dependency of travel time with respect to the number of pedestrians in the room. Solid line represents mean travel time related to given occupancy, three different pedestrians are marked to visualize the differences between participants.}
\label{fig:TT_pocet}
\end{figure}

In Figure \ref{fig:TT_pocet}, the travel time of each passing of each pedestrian is plotted against the average number of pedestrians during the interval spend by the pedestrian in the room. Of course, the higher the occupancy is, the longer is the travel time. Nevertheless, each pedestrian reacts on the high occupation (connected to larger crowd in front of the exit) differently. In Figure~\ref{fig:TT_pocet} the travel times of three different pedestrians are marked by circles: green pedestrian can be considered as aggressive one, because his travel time is significantly below average. Red circles belong to cautions non-conflict person, who rather waited at the border of the crowd. Violet circles mark an average person. The average travel time in dependence of the number of pedestrians in the room plotted as blue line.

To compare travel time measured under different conditions, the relative time (referred to as $T_R$) is presented. For every number of pedestrians inside the room $N$, average travel time $\overline{T}(N)$ is evaluated. Then, the relative time is defined as the ratio of measured time (which corresponds to given occupancy $N$) to the average travel time $\overline{T}(N)$
\begin{equation}
	T_R = \frac{T}{\overline{T}(N)}\,.
\end{equation}

\begin{figure}[h!]
	\begin{center}
	\includegraphics[width=.9\textwidth]{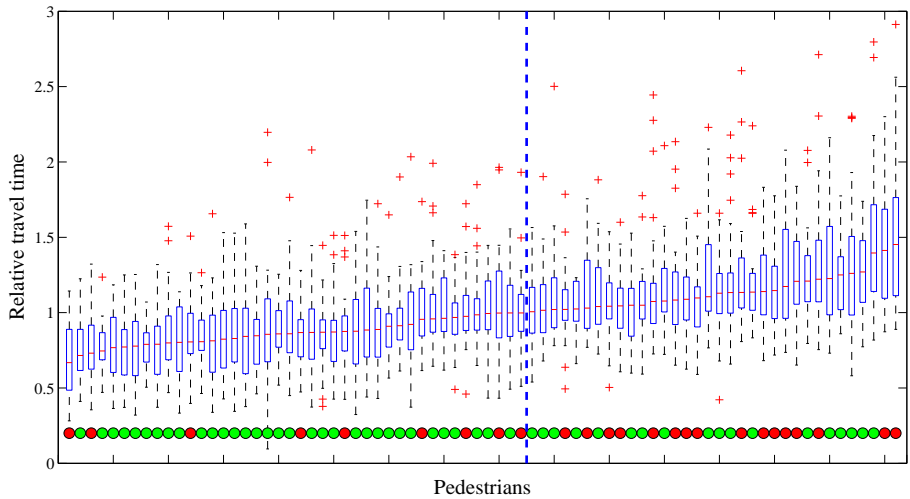}
	\end{center}
\caption{The boxplot of relative travel time, data were assigned to pedestrian and ordered with respect to median time. Color mark referes to gender (male: green, female: red).}
\label{fig:RT_box}
\end{figure}

Relative travel time can be used to analyse the differences between pedestrians in detail. In Figure~\ref{fig:RT_box}, $T_R$ is assigned to pedestrians and boxplot is provided. The records are ordered by pedestrian's median relative travel time and the gender is also visualized. It is obvious that majority of men are faster than a majority of women.

\begin{figure}[h!]
	\begin{center}
	\includegraphics[width=.45\textwidth]{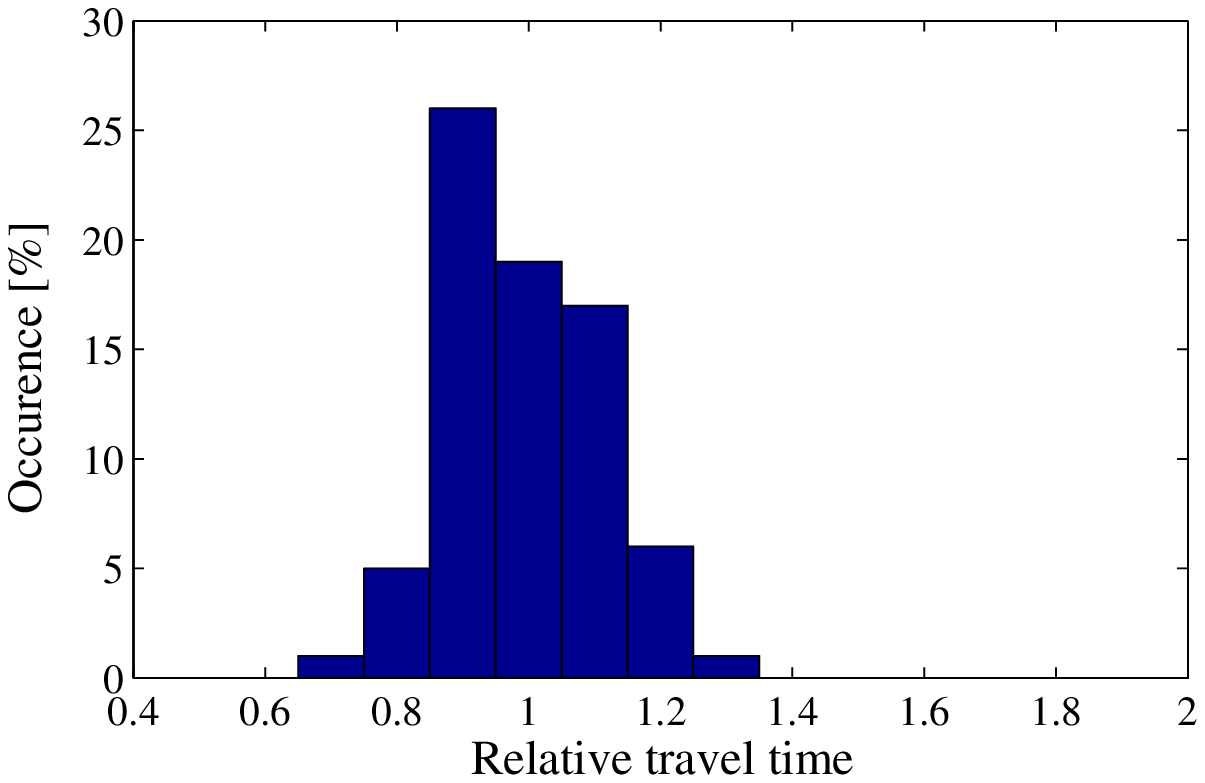}
	\includegraphics[width=.45\textwidth]{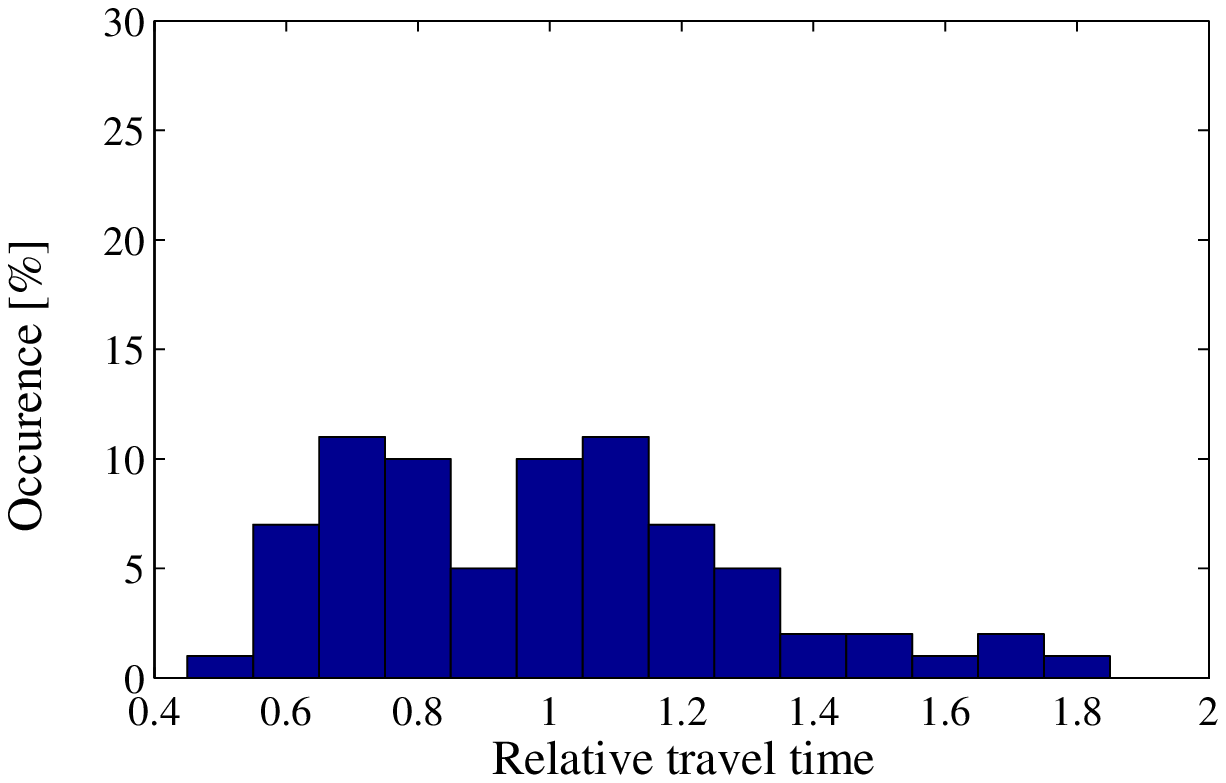}
	\end{center}
\caption{The histograms of pedestrians' median relative travel time. Left: free flow, Right: congested state.}
\label{fig:RT_hist}
\end{figure}

The differences between pedestrians increased with increasing occupation. In free flow, the travel time is determined mainly by pedestrian's desired velocity, but is the congested state the aggression plays significant role. In Figure~\ref{fig:RT_hist}, the histograms of pedestrian median relative travel time within free flow and congested state are plotted. While in free flow the slowest and fastest pedestrian differ from the mean value only by 30 \%, during the congestion median $T_R$ of slower pedestrians is several times higher. 

A related and important aspect of the evacuation experiment have been observed in the period from the stop of the inflow until the emptying of the room.  To describe the process of melting the cluster in front of the exit, the time development of pedestrian time-headways at the door (i.e. the interval between two subsequent egress times) was evaluated. 

\begin{figure}[h!]
	\begin{center}
	\includegraphics[width=.45\textwidth]{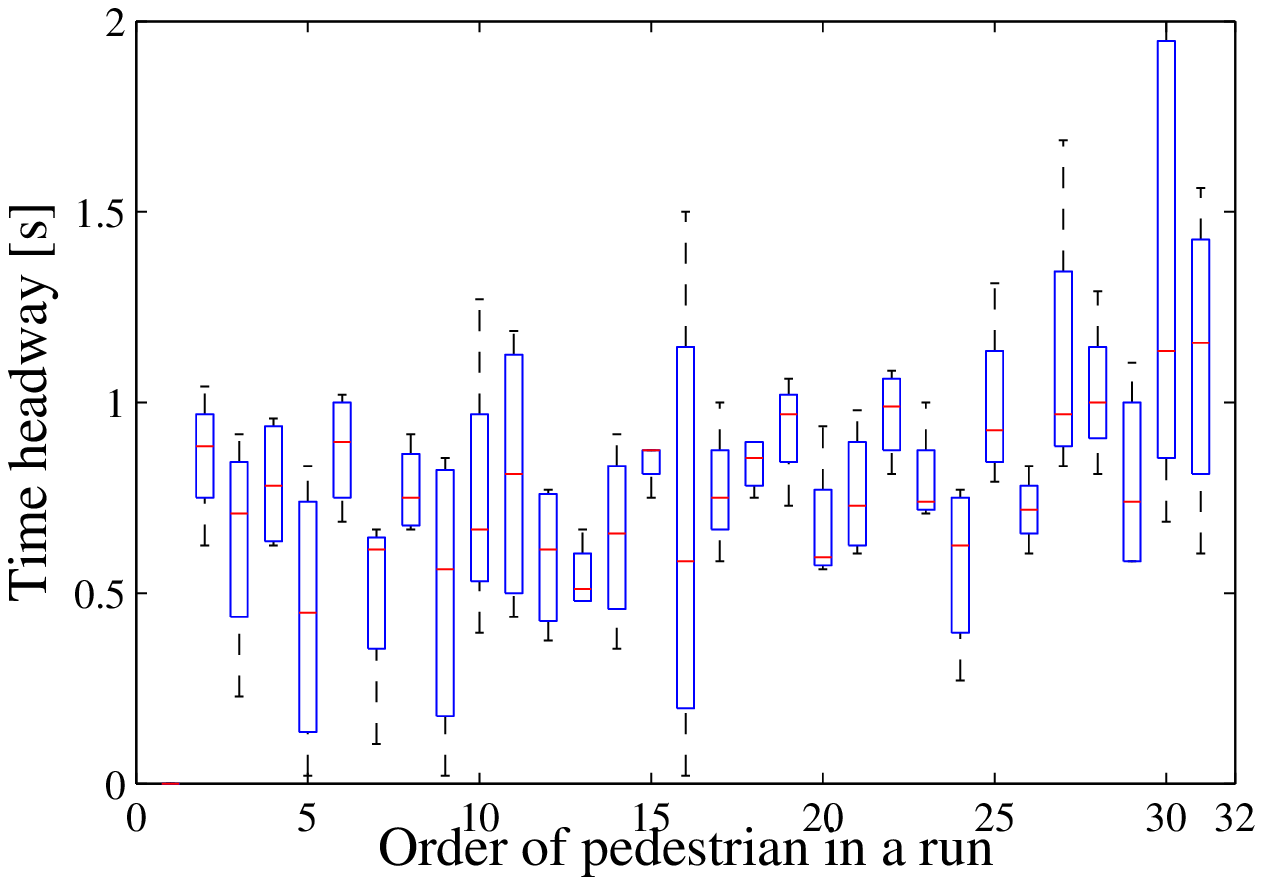}
	\includegraphics[width=.45\textwidth]{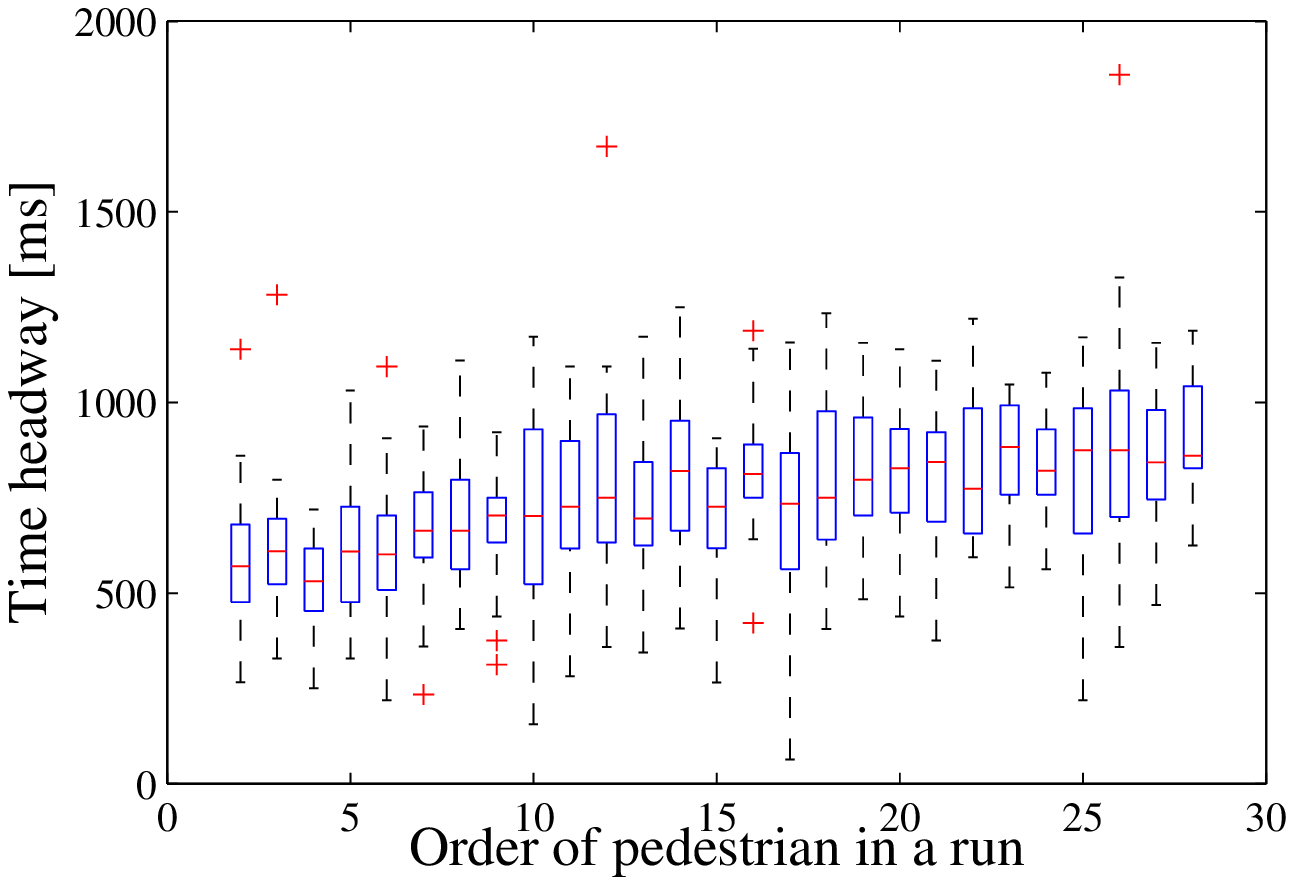}
	\end{center}
\caption{The process of melting the clusters: boxplot of time haedways of last 30 pedestrians from run \#6 -- \#10, E4.  (Left). The boxplot of time haedways form experiment leave the room E1 (Right).}
\label{fig:Head}
\end{figure}

During all experiments, the loss of motivation of last pedestrians was observed. As shown in Figure~\ref{fig:Head}, several last headways indicated increasing trend. In case of E1 (leave-the-room experiment), 30 pedestrians entered the room at the same time, therefore faster pedestrians with shorter reaction time reached the exit first and less motivated pedestrian left at the end. On the other hand in passing through scenarios, the arrivals of slow and fast pedestrians are random, thus the increasing trend of the headways is observed ``later'' when the pedestrians waiting at the edge of the crowd got to the door.

\section{Conclusions}
\label{sec:con}

From the macroscopic and microscopic observation of the experiments E2, E3, and E4, we may conclude that:
\begin{enumerate}
	\item The phase transition from the free flow to the congestion is closely related to the inflow $\alpha$ and is very sensitive on the change of the inflow value.
	\item The saturation of the room seems to be quite sudden, i.e., the outflow increases linearly with the increasing inflow. In the transition regime, the conflicts cause a slight decrease from the linear trend and after the critical value of the inflow $\alpha_{S}\approx1.4$~ped/s the outflow levels around the value 1.38~ped/s (the width of the exit is 60~cm).
	\item A good identifier of the phase transition seems to be the average travel time and the average occupation of the room. The congestion is related to the dramatic increase of pedestrians in the room followed by the increase in the average travel time.
	\item Even in the congested state, aggressive pedestrians can reach a relatively short travel-time by walking around the crowd or pushing through the crowd. On the other hand, the cautious pedestrians are not willing to fight in the crowd and their travel times are therefore significantly higher.
\end{enumerate}

Here we note that during the passing-through experiments, the volunteers have been much better motivated to leave the room and fight in the crowd then during the egress experiment E1. Therefore, such situation can better simulate the panic-like conditions. In the high density of the crowd, slight indication of the real panic has been observed, mainly caused by the fragile female students, who were trapped inside the cluster in front of the door (no one got injured during the experiment).

Thanks to the unique identification of the individual pedestrians we are able to extract several properties from the evaluation of the travel time. Concerning the    travel time in the free flow conditions, the travel time is closely related to the desired velocity of pedestrians. Considering contrarily the travel time in high density, while the number of pedestrians in the room is above 30, the travel time reflects the aggressive behaviour of the pedestrians. The identification of individual pedestrians allows even to find out the cross-correlation of those two characteristics, which can be used for the parameter distribution in pedestrian models. Such parameter analyses is the future goal of our working group.

\section*{Acknowledgement} This work k was supported by the grant SGS12/197/OHK4/3T/14. We would like to thank to second year students of FNSPE for assistance in mentioned experiments.

\bibliographystyle{plain}
\bibliography{references}

\end{document}